# What is the best firm size to invest?

Ivan O. Kitov

Institute for the Geospheres' Dynamics, Russian Academy of Sciences, ikitov@mail.ru

*Abstract*
Significant differences in the evolution of firm size distribution for various industries in the United States have been revealed and documented. For theoretical considerations, this finding puts major constraints on the modelling of firm growth. For practical purposes, the observed differences create a solid basis for selective investment strategies.

**Introduction**

This is a banal observation that firms have different sizes. It would not go far beyond this fact if the distribution of sizes is not characterized by a simple functional form, e.g. the power law, which is also common for numerous objects in physics and other natural sciences. The similarity of frequency distribution of sizes can be considered as strong evidence in favor of the existence of universal intrinsic mechanisms giving birth to the distributions. Therefore, the investigation into the processes behind the observed distribution of firm sizes is supported by the whole strength of the natural sciences. On the other hand, some specific features of various size distributions in economics and their evolution over time are likely to be helpful for the hard sciences.

There are two principal topics related to the study of firm sizes. Many researches are focused on the properties of the Pareto distribution of firm sizes; Axtell (2001), Kaizoji *et al.* (2006), Stanley *et al.* (1995) are among others. Modeling of the processes of firm growth matching the Pareto law is another area of active research (Coad, 2009; de Wit, 2005; Sutton, 1997). In general, results of the previous studies allow building a promising new branch of economics with very tight links to the hard sciences.

The purpose of our study is potentially related to both principal topics. We are trying to describe the dynamics of firm sizes distribution (FSD) in the United States, both overall and at the level of industries. Results of economic censuses in the United States are available in electronic format since 1992. Therefore, one might expect to document some measurable changes. The FSD dynamics could be helpful for the modeling of firm growth, especially if the industries under study evolve in different directions. This could evidence the existence of some inhomogeneous internal structure, which should be explained by any model of firm growth.

For the effectiveness of financial markets, it would be useful to evaluate the dependence of labor unit productivity or labor unit efficiency on firm size. Moreover, if these characteristics evolve with time, one might be able to design a sound investment strategy.

A recognized source of certified and detailed information on a variety of firms' characteristics in the United States is the Bureau of the Census, which conducts Economic Census every five years (http://www.census.gov/econ/census02/). (According to the Census Bureau documentation, there is an important difference between terms "firm" and "establishment": *Establishment* - A single physical location where business is conducted, or where services are performed; *Firm* - A business organization or entity consisting of one or more domestic establishment locations under common ownership or control. In this paper, we examine the size distribution of firms or business organizations.) Because the results of the 2007 census will be published during 2009 and 2010 we are restricted to three censuses between 1992 and 2002, as the only available in electronic format from the CB web-site. Comprehensive reports for the censuses conducted before 1992 are available in printed form.

There is a problem with compatibility of data before and after 1997 because of the change from Standard Industrial Classification (SIC) to North American Industry Classification System (NAICS). According to the Census Bureau (http://www.census.gov/econ/census02/guide/index.html): "Changes between 1997 and 2002 affect only the construction, wholesale trade, retail trade and information sectors." Hence, it is likely that these industries demonstrate some artificial changes in relevant FSDs.

The methodology of economic censuses states that those large and medium-size firms and those firms, which operate more than one establishment, have to fill a questionnaire. The size of the smallest firms is taken from relevant administrative records. Therefore, the data on the frequency distribution of firm sizes are prone to many sampling and non-sampling errors, which might result in somewhat biased conclusion.

### 1. Firm Size Distribution

First, one has to define the frequency distribution of firm size, i.e. define the measure of firm size and corresponding intervals for bin counting. There are two general approaches to the size definition – total sales of a firm as expressed in monetary units and the number of employees. Due to the inhomogeneous character of the economic censuses, one or both measures may not be available for selected industries. This also reduces the comprehensiveness of our analysis.

Figure 1 compares overall FSDs for 2002, as obtained using total sales and the number of employees. Essentially, these are the same firms counted in different bins, but their total number under investigation might not be the same for the two definitions. We exclude from calculations all firms without employees, which may have some sales, however. Also excluded are all firms in the open-end bin "… more than …". The boundary of 10,000 employees may not coincide with the $2,5000,00,000 threshold. In our study, a firm included in the original statistics may be counted either under both definitions (most common case) or under one of two definitions or is excluded.

Both FSDs in Figure 1 are normalized to the total number of firms (4,927,805 for the number of employees and 5,696,868 for total sales) and the widths of corresponding measurement bins. As a result, both curves in the Figure represent density with the following units of measurements: the portion of the total number of firms per 1K$, and the portion of the total number of firms per 1 employee, respectively. The former definition provides a better resolution for small firms and one can clearly observe the transition from quasi-exponential to power law distribution. The number of employees provides a slightly better coverage at larger

sizes. For the purposes of illustration and regression analysis, the estimates of density are assigned to average firm size, which is also reported for all predefined bins. An alternative is to associate the readings with centers of the bins or with the theoretical points, to which the density would belong according to the Pareto law. In all cases, a slight bias in the OLS regression would be observed and for the sake of the balance between simplicity and accuracy we have chosen the average firm size.

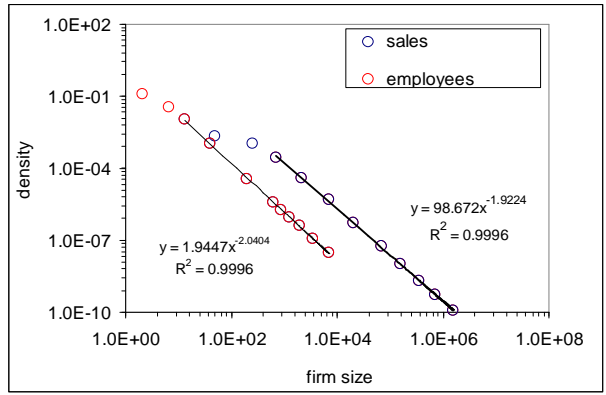

Figure 1. Two definitions of firm size: total sales in K$ and the number of employees. Both frequency distributions of size for 2002 are normalized to the total number of firms and to the widths of corresponding measurement bins. The former definition provides better resolution for small firms. Both distributions give exponents close to -2.0 consistent with $k=1$ in the Pareto law.

For larger sizes, both distributions in Figure 1 are characterized by the Pareto law (power law). Theoretically, the cumulative distribution function (*CDF*) of a Pareto distribution is defined by the following relationship:

$$CDF(x) = 1 - (x_m/x)^k$$

for all $x>x_m$, where $k$ is the Pareto index. Then, probability density function (*pdf*) is defined as

$$pdf(x) = kx_m^k/x^{k+1} \qquad (1)$$

We have estimated the Pareto index $k$ for FSDs in Figure 1. Both exponents (slopes in the log-log coordinates) are close to 2.0, and according to (1), $k=1$. These estimates do not differ from those obtained in other studies. Now we are ready to examine the dynamics of firm size distribution in the US between 1992 and 2002.

## 2. Dynamics of FSD

For 1992, the overall FSD is available only for the number of employees, and for 1997 – only for total sales. In 2002, both representations are available and supported by a smaller economic survey conduced in 2004. For better resolution of the underlying dynamics, we have naturally chosen the longest possible period between 1992 and 2002. Figure 2 displays corresponding density curves illustrating the evolution of the FSD in the US. Density for small firms practically did not change. However, there are visible discrepancies between 1992 and 2002 for mid- (the 1992 curve is above that for 2002) and large-size (the 2002 curve is above that for 1992) firms.

One has to bear in mind the transition from SIC to NAICS, however. It might add some definitional bias in the FSDs: our statistics include ~4,200,000 firms in 1992 and ~4,920,000 in 2002, for firms with more than 0 employees and less than 10,000 employees. Some firms have no employees and the bin above 10,000 employees is an open-ended one. In both cases, density can not be estimated. It is worth noting that the statistics of firm sizes measured in K$ is richer: ~4,600,000 firms in 1997 and ~5,700,000 in 2002. All in all, there were some changes in the FSD between 1992 and 2002, which might manifest themselves in different measurable features.

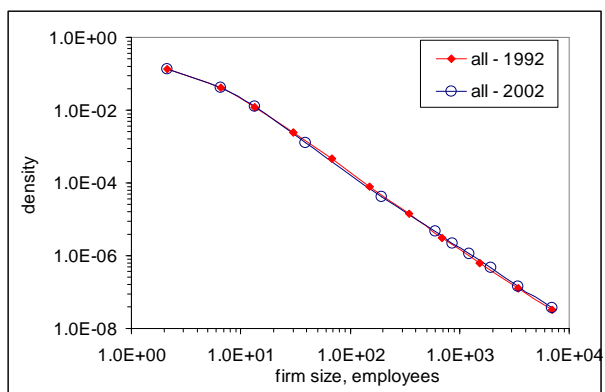

Figure 2. Dynamics of the overall FSD between 1992 and 2002. The might be just slightest discrepancies between 1992 and 2002 for mid- and large-size firms. Otherwise, the curves are indistinguishable.

An important characteristic of a firm is the efficiency of labor, as expressed by the ratio of total sales and payroll (a proxy to labor expenditures). The larger is the ratio the higher is the efficiency. In Figure 3 (left panel) we show the dependence of the efficiency on firm size. Surprisingly, the curves for 1992 and 2002 both have minimum at some size near 10 employees. Small and large firms demonstrate an elevated efficiency. The curve for 2002 is below that for 1992 for sizes smaller than 2500 employees. At larger sizes, the 2002 curve is above its counterpart. The right panel in Figure 3 illustrates the evolution of labor efficiency as a function of size – it decreases for small- and middle-size firms and increases for the largest firms. The increase reached 20% during the decade between 1992 and 2002. It would be instructive to compare the change observed in the end of the 20$^{th}$ century to that occurred in the first decade of the 21$^{st}$ century. If the trend is retained in both rate and direction, one might use it as a basis for long-term investments.

There is one obvious consequence of the labor efficiency increasing with firm size that is also observed in Figure 2. During the last decade on the 20th century, a wiser investment was in bigger firms with higher efficiency. As a result, one observed firm size redistribution and an increase in relative number of firms larger than 1000 employees. Not surprisingly, the red line (1992) in Figure 2 lies below the blue line (2002) for sizes above 1000 employees. Slightly lower labor efficiency for the firms between 10 and 1000 employees is expressed in a density decrease between 1992 and 2002. Hence, there is no contradiction between the processes in Figures 2 and 3.

This result is obtained for the overall FSD, however, and it would be of theoretical and practical importance to learn the behavior of smaller parts of the economy. Of special interest are the cases of different or even opposite dependence of labor efficiency on firm size. In other words, is the US economy homogenous or inhomogeneous in terms of the evolution of frequency distribution of firm sizes? Inhomogeneous structure of an economy would need more elaborated models of firm growth.

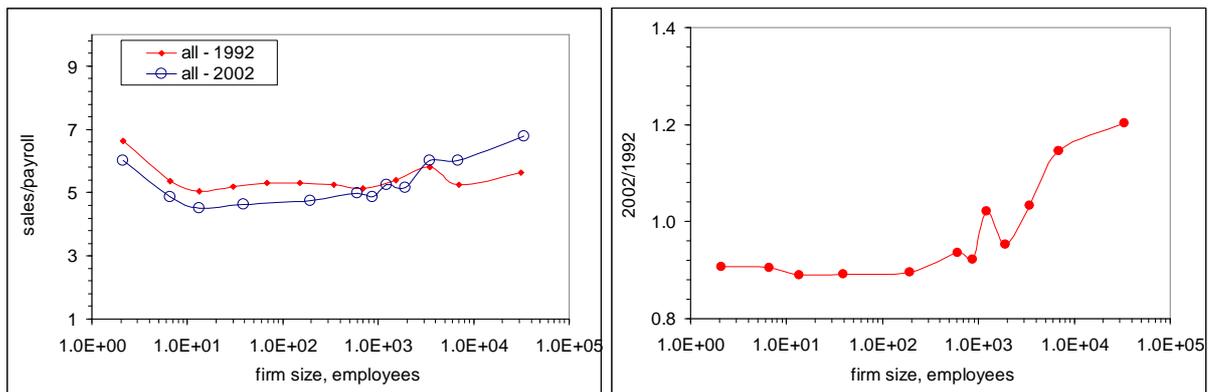

Figure 3. Labor unit efficiency (sales/payroll) as a function of firm size (left panel) and its evolution over time (right panel). Overall, larger firms provide an elevated rate of productivity growth.

Similar pattern is observed in the evolution of total sales per employee as a function of firm size. Figure 4 presents the dependence for both economic censuses and the 2002/1992 ratio. Larger firms demonstrate higher output per employee, which also increases with time. Some fluctuations near the size of 1000 employees are likely related to measurement errors.

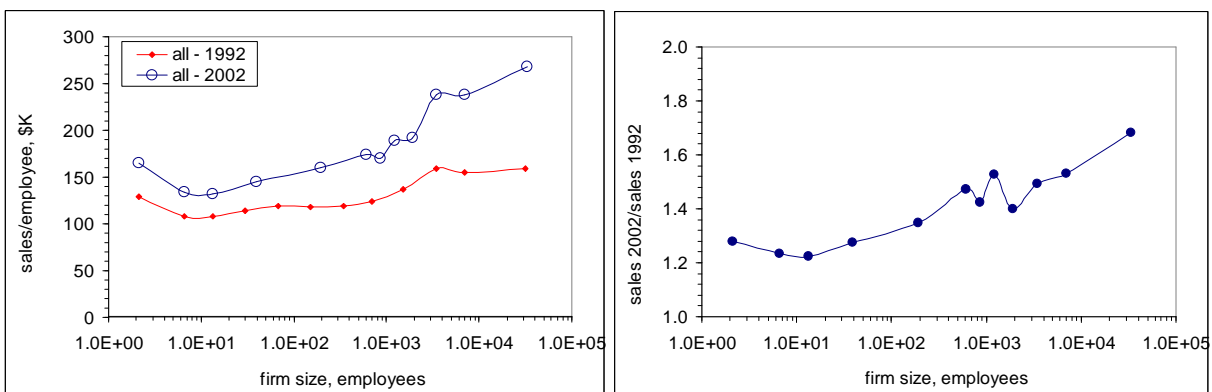

Figure 4. The evolution of total sales per employee as a function of firm size.

Having studied the overall FSD and its evolution, we now present similar curves for several selected industries. Retail is the first to analyze and likely the biggest one. The number of firms, as defined by employees, in retail decreased from ~887,000 in 1992 to 615,000 in 1997, and finally to 599,000 in 2002. Same figures are observed for the number of firms with size expressed in K$. Figure 5 displays density curves for both definitions. The left panel reveals some problems with the enumeration of firms in K$. The curve for 2002 is far above those for 1992 and 1997. The distributions over the number of employees do not show such dramatic changes. Thus, one could suggest that the discrepancy in the left panel if of artificial character.

In the right panel of Figure 5, the curves for 1992 and 2002 diverge over the whole length of the distribution. At lower sizes, the 1992 curve is below that for 2002, and in the mid-section and at the largest sizes the former curve is above the latter one. This observation is different from the pattern in Figure 2. However, the explanation of the behavior of the retail curves is similar to that given for the overall FSD – the dependence of labor efficiency on firm size and time. Figure 6 evidences that larger distances between the 1992 and 2002 curves correspond to relative decrease in density in Figure 5, and vice versa. For example, the largest distance is observed for sizes between 100 and 1000 employees, where the 1992 FSD curve is clearly above the 2002 curve. This observation supports the mechanism of predominant investment in more labor effective firms, as discussed above.

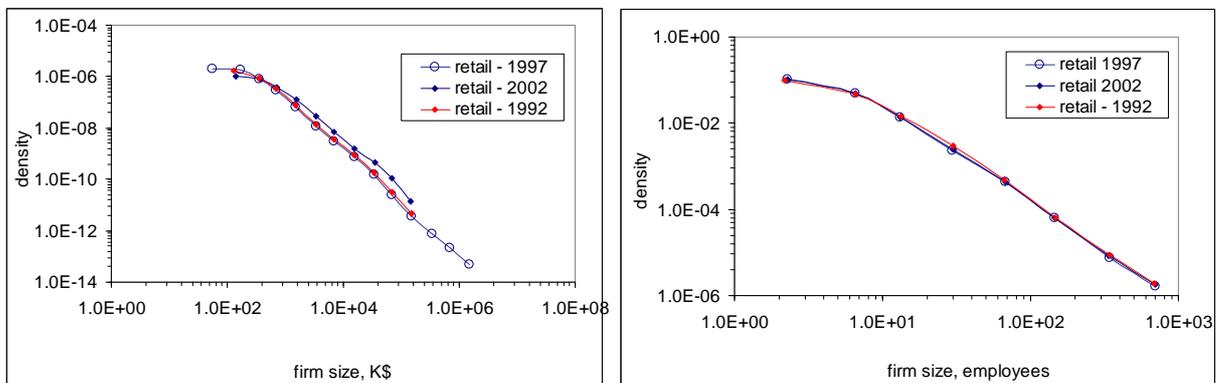

Figure 5. Dynamics of the FSD for retail between 1992 and 2002.

There is a very specific feature characterizing the curves in Figure 6 – a deep trough for sizes between 100 and 250 employees. The trough also slightly deepened relative to the peak, which jumped from the smallest firms in 1992 to those near 1000 employees in 2002. This implies a relative decrease in labor efficiency for the mid-size firms. To develop a reliable investment strategy in the segment of retail, one should keep track of the observed tendency. In any case, the decreasing density of the retail FSD in the range between 50 and 250 employees implies a higher probability of such firms to shrink or even to fail.

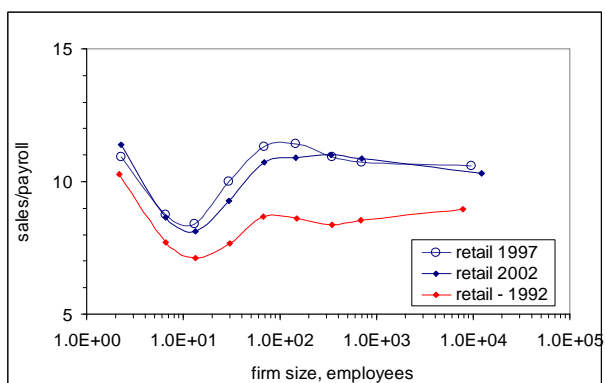

Figure 6. Retail: labor efficiency (sales/payroll) as a function of firm size and its evolution over time.

Wholesale is the next industry to present. Corresponding FSDs are depicted in Figure 7. In general, the evolution is similar to that observed for the overall FSD – relative decrease in density for mid-size firms and increase – for the larger firms. The effect is more prominent for the wholesale firms. However, with ~436,000 firms in 1992 and ~377,000 in 2002, the influence of the wholesale on the overall FSD is weak.

The sales/payroll ratio in Figure 8 demonstrates a robust increase with firm size. Hence, bigger firms are characterized by higher labor efficiency, which induces the observed increase in density of the FSD at sizes above $2,500,000 (left panel). It is important that the level of the ratio uniformly decreases with time for all sizes. Unfortunately, there are no data on total sales and payroll for 1992, which could help to better resolve the evolution.

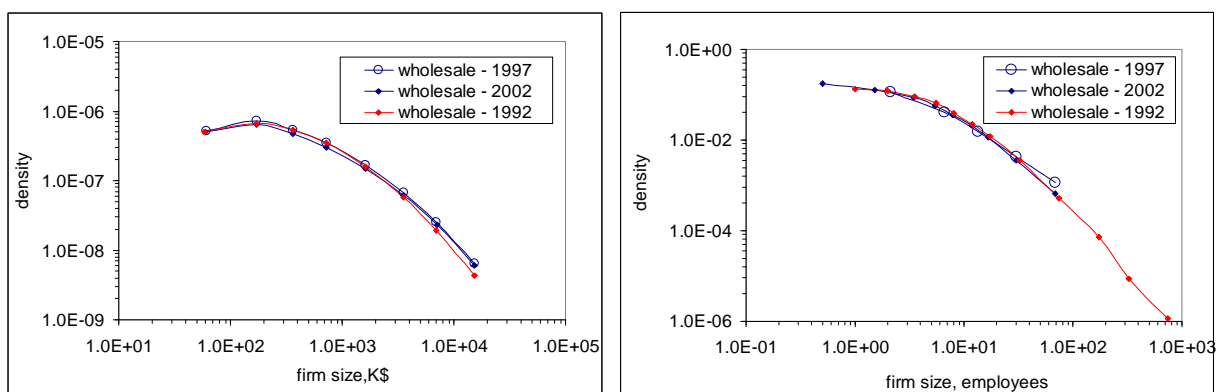

Figure 7. Dynamics of the FSD for wholesale between 1992 and 2002.

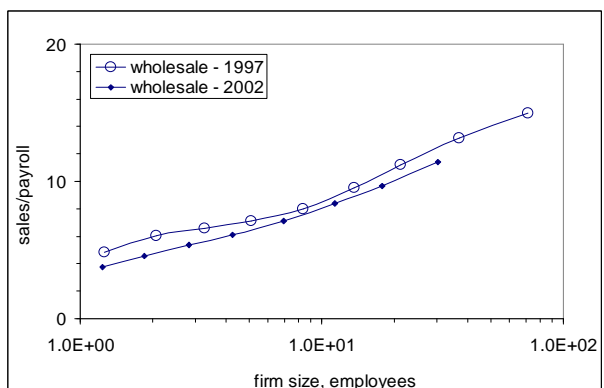

Figure 8. Labor unit efficiency (sales/payroll) as a function of firm size and its evolution over time.

Manufacturing reveals a more stable FSD – the curves for 1992, 1997, and 2002 in Figure 9 almost coincide over the entire range with only slight deviations for very small and the largest sizes. Labor efficiency demonstrates a plateau at small sizes and then increases with size. The ratio also shows a weak tendency to decrease over time. For very large sizes, statistics is not reliable with only ~1000 firms having from 1,000 to 2,500 employees.

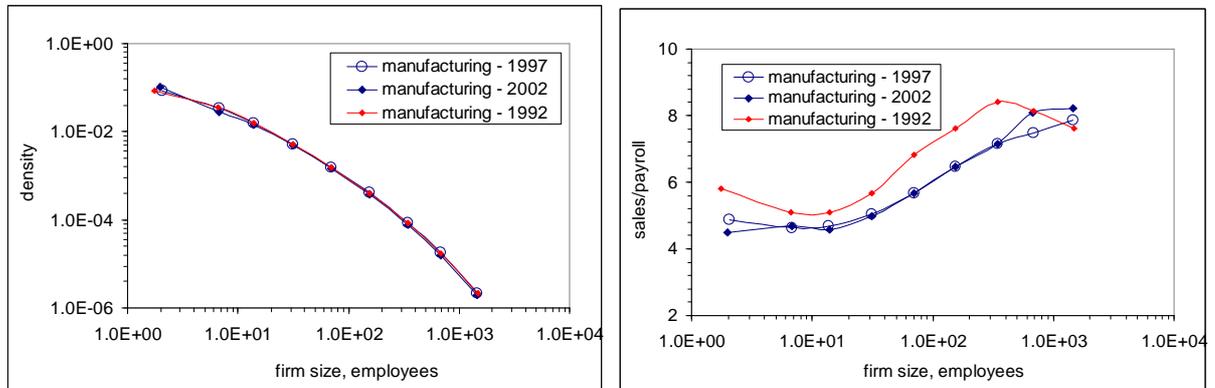

Figure 9. **Left panel**: dynamics of the FSD for manufacturing between 1992 and 2002. **Right panel**: labor unit productivity (sales/payroll) as a function of firm size and its evolution over time.

The FSDs for construction displayed in Figure 10 reveal a very clear evolutionary picture – density at larger sizes monotonically increases over time. This observation is partly supported by the increase in labor efficiency at larger sizes, as depicted in Figure 11. Another remarkable feature demonstrated by the sales/payroll ratio for the construction industry is the efficiency diminishing with firm size.

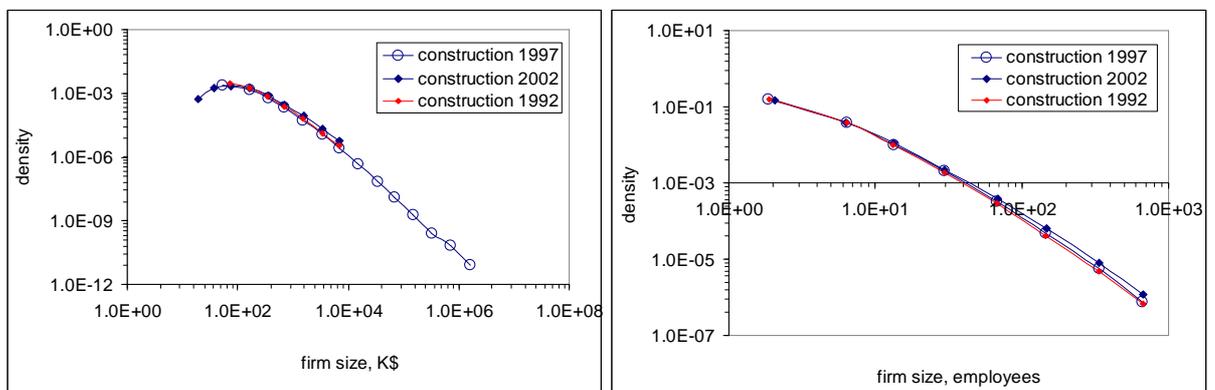

Figure 10. Dynamics of the FSD for construction between 1992 and 2002.

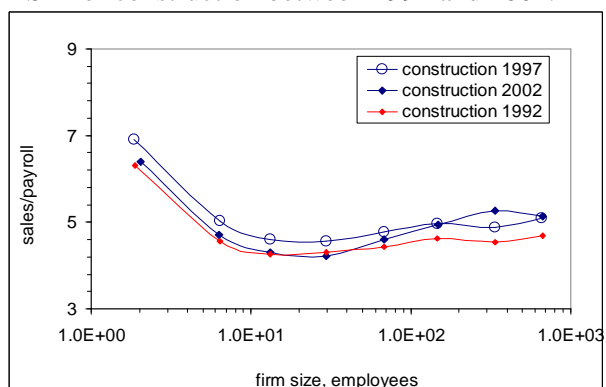

Figure 11. Labor unit efficiency (sales/payroll) as a function of firm size and its evolution over time

Mining industry presented in Figure 12 is characterized by a very stable FSD. The sales/payroll ratio is similar to that of the manufacturing – a general decrease in labor efficiency with size, but the effect is less prominent. In this regard, construction and mining are unique cases among all industries considered in the study. There are some other industries to examine, however.

Any tactics or strategy for those who want to invest in mining should take into account the discouraging behavior of labor efficiency. The returns from biggest companies will likely not be growing with time. On the contrary, in relative terms, the biggest companies suffer a higher failure rate than the smallest ones.

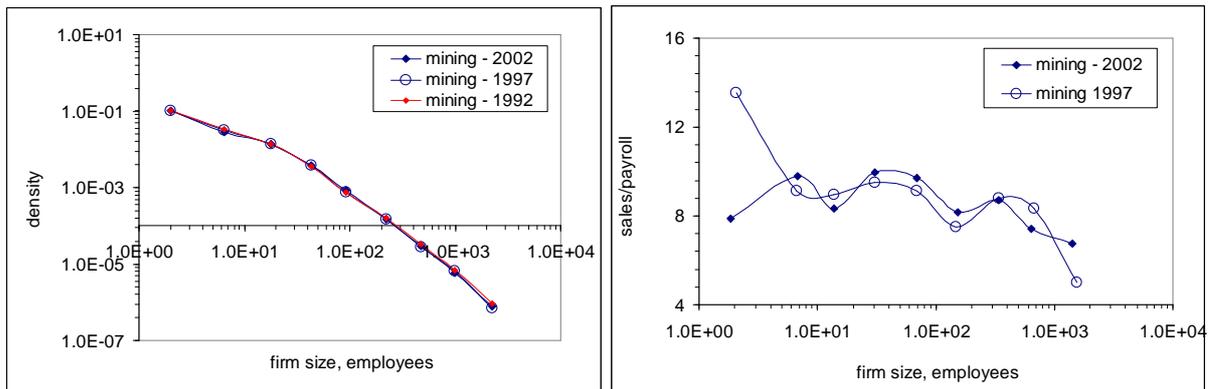

Figure 12. **Left panel**: dynamics of the FSD for mining between 1992 and 2002. **Right panel**: labor unit efficiency (sales/payroll) as a function of firm size and its evolution over time

**Conclusion**

We have documented different types of the evolution of FSD as demonstrated by various industries in the United States. For theoretical purposes, one has to bear in mind the entire diversity when building a model of firm growth. There should be one or several factors which define the observed behavior. The sales/payroll ratio is a good candidate for the explanation of the overall pattern, but its own behavior has to be modelled as well. In any case, the US economy is inhomogeneous in terms of firm size distribution, when it is decomposed in several sectors according to industrial classification system. Despite the flavor of artificiality of such decomposition for theoretical consideration, the inhomogeneous structure waits for a quantitative description.

For practical purposes, these inhomogenieties and the dependence of labor efficiency on firm size provide a reliable and fruitful basis for the development of long-term investment strategies. Those robust trends, which were observed in the size dependencies of the sales/payroll and sales/employee ratios for the studied industries between 1992 and 2002, when and if confirmed by the 2007 census, would be the first profitable candidates. One can easily choose appropriate industry and optimal firm size to the best long-term investment. The strategy might be enhanced by a sound choice of an industry, which provides the highest rate of price growth relative to other sectors (Kitov and Kitov, 2008; Kitov, 2009)

## References

Axtell, R. , (2001). Zipf distribution of US firm sizes. Science, 293:1818–1820.

Coad, A., (2009). The Growth of Firms: A Survey of Theories and Empirical Evidence. Edward Elgar: Cheltenham, UK.

de Wit, G. (2005). Firm size distributions an overview of steady-state distributions resulting from firm dynamics models. International Journal of Industrial Organization, 23(5-6):423–450.

Kaizoji, T., Iyetomi, H., Ikeda, Y., (2006). Re-examination of the Size Distribution of Firms, Evolutionary and Institutional Economics Review, 2 183-198.

Kitov, I., (2009). Apples and oranges: relative growth rate of consumer price indices," MPRA Paper 13587, University Library of Munich, Germany.

Kitov, I., Kitov, O., (2008). Long-Term Linear Trends In Consumer Price Indices, Journal of Applied Economic Sciences, Spiru Haret University, Faculty of Financial Management and Accounting Craiova, vol. 3(2(4)_Summ).

Stanley, M., Buldyrev, S., Havlin, S., *et al.,* (1995). Zipf plots and the Size Distribution of Firms, Economics Letters. 49 453-457.

Sutton, J. (1997). Gibrat's legacy. Journal of Economic Literature, XXXV:40–59.